\documentclass[a4paper,twocolumn,showpacs,showkeys]{revtex4}
\usepackage{amsmath}
\usepackage{amsfonts}
\usepackage{amssymb}
\usepackage[pagebackref]{hyperref}
\usepackage{bm}
\usepackage{graphicx}

\begin{document}

\title{Rotating inclined cylinder and the effect of the tilt angle on vortices}

\author{R.~H\"anninen}
\affiliation{Low Temperature Laboratory, Helsinki University of Technology, \\ P.O.Box
5100, FI-02015 TKK, Finland \\
\email{Risto.Hanninen@tkk.fi}}



\date{\today}



\begin{abstract}

We study numerically some possible vortex configurations in a rotating cylinder that is
tilted with respect to the rotation axis and where different numbers of vortices can
be present at given rotation velocity. In a long cylinder at small tilt angles the vortices 
tend to align along the cylinder axis and not along the rotation axis. We also show that 
the axial flow along the cylinder axis, caused by the tilt, will result in the 
Ostermeier-Glaberson instability above some critical tilt angle. When the vortices become 
unstable the final state often appears to be a dynamical steady state, which may contain 
turbulent regions where new vortices are constantly created. These new vortices push other 
vortices in regions with laminar flow towards the top and bottom ends of the cylinder where they 
finally annihilate. Experimentally the inclined cylinder could be a convenient 
environment to create long lasting turbulence with a polarization which can be adjusted
with the tilt angle.    

\end{abstract}

\maketitle

\section{Introduction}\label{s.intro}

Under uniform rotation the equilibrium state of a superfluid consists of a 
regular array of rectilinear quantized vortices along the rotation axis. 
This is strictly true only when the top and bottom plates of the rotating 
container are perpendicular to the rotation axis and its cross section has 
fixed translationally invariant form along the rotation axis. Near the walls
that are parallel to the rotation axis the vortex array is surrounded by a small vortex 
free layer of thickness that is of the order of the vortex spacing\cite{fetterSurf,donnellySurf}.

The usual setup to study vortices in rotation is to use a cylindrical
container and to assume that the cylinder axis is aligned exactly parallel
to the rotation axis. Nevertheless, experimentally the rotation axis is
always tilted by some small unknown angle owing to uncontrolled mechanical
misalignment. Even if the misalignment is small, one is tempted to ask the
question: What is the effect of the tilt and how does it affect the steady
state vortex configuration? Fortunately, it turns out that cylindrical
symmetry is not immediately broken at any arbitrarily small tilt angle.
A sensitive test of the tilt angle is the measurement of the number and configuration
of vortices in the equilibrium state of rotation\cite{Ruutu1998}, particularly
the relative number of the centrally located rectilinear vortices and the curved 
peripheral vortices which have one or both ends on the cylindrical side wall.
During changes of the rotation velocity $\Omega$ the curved peripheral vortices
behave very differently from the central rectilinear vortices: They have no
annihilation energy barrier\cite{Ruutu1998} during a deceleration of $\Omega$, 
while during acceleration they may become unstable with respect to loop
formation and reconnection at the cylindrical side wall which leads to the 
generation of new independently evolving vortices in the increased rotational 
flow\cite{precursorPRL}.

For simple flow geometries at low velocities vortex filament calculations are 
today a reasonably efficient and reliable way to find the resulting vortex
configurations, in many cases with less effort than by direct measurement.
The rotating inclined cylinder is a good example since an extensive series
of measurements would be needed to scan the behavior as a function of tilt
angle, rotation velocity, and mutual friction. Here we are not attempting
to provide the entire quantitative picture of the possible flow states,
but look for the main qualitative changes from the ideal case.

Already in the 1980's Mathieu {\it et al.}\cite{mathieu} considered a cavity 
that was tilted with respect to the rotation axis. They used the free-energy 
principle combined with the HBVK continuum approximation.
They also implicitly assumed that far from the boundaries the vortices are aligned 
along the rotation axis, and considered a rectangular cavity which is infinite 
in one direction (that is perpendicular to the rotation axis). 
In the rectangular geometry considered by Mathieu {\it et al.}\cite{mathieu} the superfluid
velocity without vortices is always along the direction which is considered to
to be infinitely long. This forces the vortices to lie in the plane perpendicular
to this direction. In reality the cavities are not infinitely long and often
the longest dimension is oriented along the rotation axis. Therefore the 
flow of the superfluid component without vortices takes a more complicated form. 
Unless the container has rotational symmetry and its symmetry axis is aligned perfectly 
along the rotation axis, the superfluid is not at rest any more but moves due to the container 
boundaries. This results in that the superfluid velocity is not simply caused by vortices but 
contains a potential part that must be taken into account.  

In the following we consider a circular cylinder and find its vortex configuration 
in rotation using the vortex filament model. The stable vortex configuration is 
obtained by following the evolution in the vortex dynamics long enough in order to 
obtain the steady state response. The most striking result, which differs from the 
general view and from the results assumed in Ref.~\cite{mathieu},
is that far from boundaries the vortices may tend to align not along the
rotation axis but along the symmetry axis of the rotating cylinder. This is especially 
true for a single vortex under high rotation and for an array with a large number 
of vortices. These configurations would result if the cylinder with a given number 
of straight vortices is tilted by an angle which is less than a critical value. 
With large enough tilt angles the vortices may become unstable and tend to 
orient more along the rotation axis. Above the critical angle there exist 
turbulent regions in the tilted cylinder which act as a source for new vortices. 
Therefore the tilt of the cylinder has a similar effect as rotation around
more than one axis\cite{Kobayashi}.

The article is organized as follows. First in Sec. \ref{s.idealfluid} we 
solve the velocity profile for the superfluid component inside the cylinder
when no vortices are present. In the following section we describe briefly 
the vortex filament model and the numerical procedure that is used to 
determine the vortex motions and the steady state inside the cylinder. 
Section \ref{s.results} is devoted to the results.

\section{Motion of the ideal fluid inside tilted rotating cylinder}\label{s.idealfluid}

\begin{figure}[h]
\centerline{\includegraphics[width=0.5\linewidth]{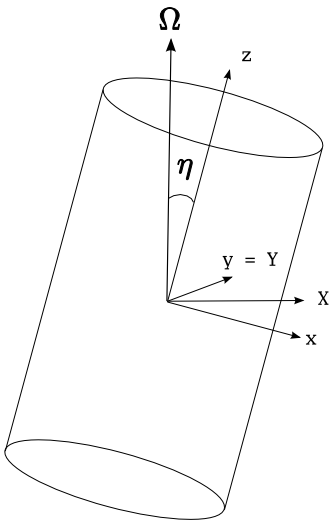}}
\caption{Cylinder tilted by an angle $\eta$ with respect to the rotation
axis $\bm{\Omega} = \Omega\hat{\bf Z}$.}
\label{f.ekakuva}
\end{figure}

The helium superfluids, $^4$He and $^3$He, can be described using a two-fluid
model where the superfluid part behaves like an ideal Eulerian fluid with negligible 
viscosity and the normal component behaves like a classical fluid with finite 
viscosity. We are mostly interested in superfluid $^3$He-B, where the normal
component is very viscous and can always be considered to be in solid
body rotation in the rotating cavity. If there are no vortices, 
the superfluid part is in potential flow, which must be taken into 
account before one can consider the motion of the vortices. In case 
of an ideal circular cylinder, with radius $R$ and length $L$, 
aligned perfectly along the rotation axis, the superfluid remains 
stationary in the laboratory frame. When the cylinder axis is not parallel 
to the rotation axis the superfluid is no more at rest but is put into motion 
due to the boundaries. The same occurs also in all other cases without cylindrical
symmetry, such as the rotating box\cite{MT} or a cylinder with ellipsoidal cross section\cite{LL}.
Consider a situation (at some particular time) illustrated in Fig.\ 
\ref{f.ekakuva} where  $\bm{\Omega} = \Omega\hat{\bf Z}$ and the cylinder 
axis is along $\cos\eta\hat{\bf Z}+\sin\eta\hat{\bf X}$. At that particular 
time the fixed Cartesian coordinates $(X,Y,Z)$, denoted by uppercase letters, 
and the Cartesian coordinates $(x,y,z)$, fixed to the rotating cylinder and denoted
by lowercase letters, can be related by a rotation by an angle $\eta$ around the $Y$-axis. 
The cylindrical coordinates $(\rho,\phi,z)$ that are also fixed to the rotating
cylinder, such that $\rho=\sqrt{x^2+y^2}$ and $\phi=\arctan(y/x)$, can then be related
to the fixed laboratory coordinates via 
\begin{eqnarray}
X &=& \rho \cos\phi \cos\eta + z \sin\eta \nonumber \\
Y &=& \rho \sin\phi \\
Z &=& - \rho \cos\phi \sin\eta + z \cos\eta . \nonumber
\end{eqnarray}

\begin{figure}[h]
\centerline{\includegraphics[width=\linewidth]{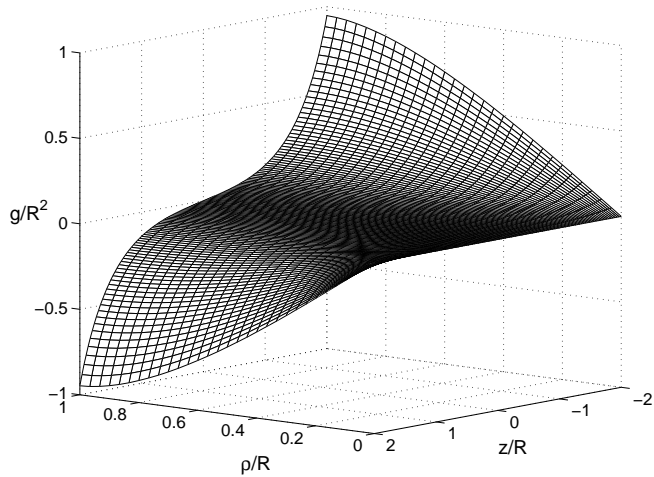}}
\caption{Solution for velocity potential $g(\rho,z)$ when $L=4R$.}
\label{f.samplegsol}
\end{figure}

\begin{figure}[h]
\centerline{\includegraphics[width=\linewidth]{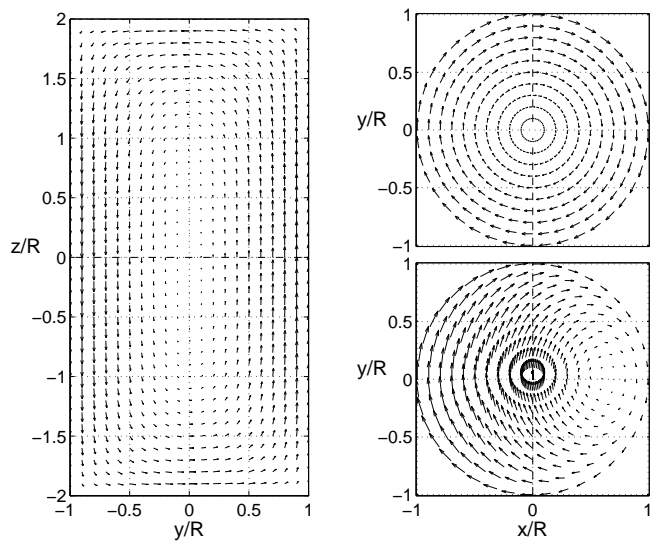}}
\caption{Velocity in the rotating cylinder when $L=4R$ and when the 
angle between the rotation axis and cylinder axis, $\eta = 30^\circ$.
{\it Left:} Velocity in the $x=0$ plane, the plane that is perpendicular to
the plane defined by the cylinder and the rotation axes. 
{\it Right top:} Velocity in the $z=0$ plane. 
{\it Right bottom:} Velocity in the bottom of the cylinder, $z=-L/2$. 
The coordinates $(x,y,z)$ are fixed to the rotating cylinder.}
\label{f.samplevelo}
\end{figure}

\noindent
The boundaries of the cylinder are given by the surfaces $\rho = R$ and $z=\pm{L}/2$. 
In cylindrical coordinates $\bm{\Omega}$ and ${\bf r}$ are given by
\begin{eqnarray}
\bm{\Omega} &=& \Omega\left[-\sin\eta \cos\phi\hat{\bm{\rho}} 
+ \sin\eta \sin\phi \hat{\bm{\phi}} + \cos\eta\hat{\bf{z}} \right] \\
{\bf r} &=& \rho \hat{\bm{\rho}} + z \hat{\bf{z}} .
\end{eqnarray}
Without vortices the superfluid velocity follows that in potential flow, 
${\bf v} = \nabla\Phi$, where the potential $\Phi$ satisfies the Laplace 
equation. The boundary condition is that at the surfaces (with normal $\hat{\bf n}$)
the velocity component normal to the boundary satisfies 
$\frac{\partial\Phi}{\partial{n}}=\hat{\bf n}\cdot(\bm{\Omega}\times{\bf r})$
which implies that one must solve the following system of equations:
\begin{eqnarray}
\label{e.diffequ1}
\frac{1}{\rho}\frac{\partial}{\partial{\rho}} \left(\rho \frac{\partial\Phi}{\partial{\rho}} \right)
+ \frac{1}{\rho^2}\frac{\partial^2\Phi}{\partial\phi^2}+\frac{\partial^2\Phi}{\partial{z}^2} &=& 0 \\
\frac{\partial\Phi}{\partial{\rho}}_{\vert_{\rho=R}} = \Omega{z} \sin\eta \sin\phi, \hspace{4mm}
\frac{\partial\Phi}{\partial{z}}_{\vert_{z=\pm\frac{L}{2}}} &=& -\Omega \rho \sin\eta \sin\phi. \nonumber
\end{eqnarray}
The above equations give the solution in the laboratory frame. Conversion
to the rotating frame is obtained by removing the effect of rotation, 
so that the velocity is given by
\begin{equation}
{\bf v} = \nabla\Phi - \bm{\Omega} \times {\bf r} . \label{e.velorot}
\end{equation}
Equation (\ref{e.diffequ1}) can be further simplified with the use of a trial function
$\Phi(\rho,\phi,z)$ $=$ $\Omega (\rho z+g(\rho,z))\sin\phi \sin\eta$,
which results in that $g(\rho,z)$ has to satisfy the following Helmholtz equation:
\begin{eqnarray}
\frac{1}{\rho}\frac{\partial}{\partial{\rho}} \left(\rho \frac{\partial{g}}{\partial{\rho}} \right)
+\frac{\partial^2{g}}{\partial{z}^2} - \frac{1}{\rho^2}g &=& 0 \nonumber \\
\frac{\partial{g}}{\partial{\rho}}_{\vert_{\rho=R}} = 0,   \hspace{5mm}
\frac{\partial{g}}{\partial{z}}_{\vert_{z=\pm\frac{L}{2}}} &=& -2\rho . \nonumber
\label{e.diffequ2}
\end{eqnarray}
Note that the potential $g(\rho,z)$ does not depend on tilt angle $\eta$.
The velocity components in the rotating frame, obtained from Eq.\ (\ref{e.velorot}), 
are then given by
\begin{eqnarray} \label{e.velorot2}
v_\rho     &=& \Omega \sin\phi \sin\eta \frac{\partial g}{\partial \rho} \nonumber \\
v_\phi  &=& \Omega \cos\phi \sin\eta \frac{g}{\rho} - \Omega \rho \cos\eta \\
v_z &=& \Omega \sin\phi \sin\eta (2\rho+\frac{\partial g}{\partial{z}}) . \nonumber
\end{eqnarray}
One notices that the tilt of the cylinder implies a flow parallel to the cylinder 
axis and also reduces the azimuthal counterflow. The axial velocity is largest at the 
cylinder wall ($\rho=R, z=0$) and varies as $\sin\phi$ as one goes around the cylinder axis. 
The solution to the above equations is found numerically, except in case of an infinitely 
long cylinder, when $g = 0$. As an example, the solution for the vortex-free flow state
is plotted in Fig.~\ref{f.samplegsol} for the potential $g(\rho,z)$ and in Fig.\ \ref{f.samplevelo}
for the velocity $v(x,y,z)$, when $L/R = 4$ and $\eta = 30^\circ$.

\section{Vortex filament model}\label{s.vfm}

Superfluid vortices are modeled as line defects. This model is suitable especially
for superfluid $^4$He where the core diameter is of the order of atomic size. Our
calculations are attempting to model primarily superfluid $^3$He-B where two simplifications
can be assumed: 1) The normal fraction is always in solid-body rotation with the container,
as noted in Sec.\ \ref{s.idealfluid}. 2) In addition the core diameter $a_0\approx 10^{-5}$ mm. This
is still a small value compared to the experimental dimensions of a few millimeters and 
the use of the vortex filament model is appropriate. The larger coresize, however,
makes it more justifiable to assume smooth walls for the cylinder, since so far 
there has been no clear experimental indication that vortices become pinned if the walls 
of the container are well polished\cite{PLTP16}. Also down to intermediate temperatures
of $T \gtrsim 0.4T_c$ mutual friction damping is sufficiently high so that vortices are
rapidly annihilated at zero flow conditions. In such conditions vortex remanence is not a
major problem and relatively large vortex-free normal fluid-superfluid counterflow can be
achieved. In practice this means that the rotating container can generally be filled with
any number of vortices up to the number which corresponds to the equilibrium state with
solid-body rotation of the superfluid component\cite{Ruutu1997}. 

The vortex filament model is well described by Schwarz in Refs.\cite{schwarz85,schwarz88}.
Our numerical scheme is similar. Line vortices are described by a sequence of points 
$\lbrace{\bf s}_i\rbrace_{i=0}^M$. The required tangent, curvature radius {\it etc}. at some 
particular point ${\bf s}({\bf r})$ along the line vortex are calculated by fitting a 
circle through three points (the point under consideration and the nearest neighbours). 
The main difference, compared to calculations of Schwarz, is that we use a 4th order 
Runge-Kutta method to determine the vortex motion that is described by the following equation:  
\begin{equation} 
{\bf v}_{\rm L}={\bf v}_{\rm s} +\alpha
\hat{\bf s}' \times ({\bf v}_{\rm n}-{\bf v}_{\rm s}) -\alpha'
\hat{\bf s}' \times [\hat{\bf s}' \times ({\bf v}_{\rm n}-{\bf v}_{\rm s})]\,.
\label{e.vl}
\end{equation}
Here ${\bf v}_{\rm L}=d{\bf s}/dt$ is the velocity of the vortex point at ${\bf s}$ with unit 
tangent $\hat{\bm{s}}'$.  The above equation includes the mutual friction between vortex 
lines and the normal fluid with coefficients $\alpha$ and $\alpha'$ that are taken as a 
function of temperature from the measurements of Bevan {\it et al.}\cite{Bevan,BevanPRL}
This equation is typically used in the laboratory frame where 
${\bf v}_{\rm n} = \bm{\Omega} \times {\bf r}$, but is also valid in the 
rotating frame where ${\bf v}_{\rm n} = 0$.\cite{sonin}
Now in the rotating frame the superfluid velocity ${\bf v}_{\rm s} = {\bf v} + {\bf v}_{\omega} 
+ {\bf v}_{\rm b}$ consists of the velocity due to rotation, ${\bf v}$, calculated 
above in Sec. \ref{s.idealfluid} and given by Eq.~(\ref{e.velorot2}), 
the superfluid velocity due to vortices, ${\bf v}_\omega$, 
and the boundary induced velocity, ${\bf v}_{\rm b}$, that is needed to cancel the velocity 
through the boundaries due to vortices. 

As one may notice the above method for doing the calculation in the rotating frame differs
from previous simulations in Refs.\cite{nature,TsubotaPRB2004,TsubotaPRL2003}, which contain 
a complicated term, $\dot{\bf s}_{\rm rot}$, that was used to take into account the rotation.
This earlier approach cannot be completely correct. A simple way to check this is to consider the 
zero temperature limit in an infinite system with no other external velocities (except the rotation 
of the normal component, which at $T = 0$ cannot couple to the vortex motion, simply because 
it is absent). In this case our ${\bf v}$ reduces to ${\bf v}=-\bm{\Omega}\times{\bf r}$ and that 
takes care of the rotating frame. This term does not change the vortex configuration, but just 
rotates the coordinates, like expected. However, the term $\dot{\bf s}_{\rm rot}$ may have a
more dramatic change on the vortex configuration since the term depends on the vortex configuration
itself. Therefore, owing to its complex form, this term has different and configuration-dependent
consequences.

The superfluid velocity due to vortices is obtained using the Biot-Savart formulation:
\begin{eqnarray}
{\bf v}_{\omega}({\bf r},t) = \frac{\kappa}{4\pi}
\int\frac{({\bf s}_1-{\bf r})\times d{\bf s}_1}
{\vert {\bf s}_1-{\bf r}\vert^3}. \label{e.vs}
\end{eqnarray}
Here the integration is a line integral along the vortices and $\kappa$ is the quantum 
of circulation. When ${\bf r}={\bf s}$ is one of the vortex points, the above equation 
becomes singular as ${\bf s}_1\rightarrow{\bf s}$. This singularity can be avoided, for example, 
by considering a moving vortex ring \cite{schwarz85}. If one denotes with $l_{-}$ and $l_{+}$ 
as being the lengths of two adjacent line elements connected to the point ${\bf s}$, 
then the velocity of this point is given by
\begin{eqnarray}\label{e.bs}
{\bf v}_{\omega}({\bf s},t) &=& \frac{\kappa}{4\pi}\hat{\bf s}'\times {\bf s}'' 
\ln\left(\frac{2\sqrt{l_{+}l_{-}}}{e^{1/4}a_0}\right) \nonumber \\
&+&\frac{\kappa}{4\pi}\int^{'}\frac{({\bf s}_1-{\bf s})\times d{\bf s}_1}
{\vert {\bf s}_1-{\bf s}\vert^3}. 
\end{eqnarray}
The integral (the term that is typically denoted as the non-local term) is now taken over 
the rest of the vortex filament, not connected to the point ${\bf s}$. Vectors 
$\hat{\bf s}'$ and  ${\bf s}''$ in the local term are the tangent and principal normal 
at ${\bf s}$ and the derivation is with respect to the arc length, $\xi$, along the vortex. 

As already described by Schwarz, the boundary velocity, ${\bf v}_{\rm b} = \nabla\Phi_{\rm b}$, 
in general, must be obtained by solving the Laplace equation $\nabla^2\Phi_{\rm b}=0$
with the boundary condition that $\hat{\bf n}\cdot({\bf v}_\omega+{\bf v}_{\rm b})=0$ 
on the boundary, with normal $\hat{\bf n}$. This method has been used in several 
previous calculations, for example in Ref.~\cite{twistPRL}, by using finite differences to
discretize the Laplace equation and solving the resulting sparse matrix equation. The 
resulting matrix is typically very large and limits the grid size that can be used to 
evaluate the boundary velocity. In these calculations here we used an approximate method,
namely image vortices to evaluate the boundary velocity. Above and below the cylinder we used
simple image vortices to cancel the velocity through the bottom and top plates of the
cylinder. The velocity through the cylindrical wall at $\rho=R$ is canceled 
using an inverse image curve where the distance of the image vortex from the cylinder axis 
is given by $\rho_{\rm im} = R^2/\rho$. Here $\rho$ is the distance of the vortex from 
the axis. The $z$ and $\phi$ coordinates of the image vortex are the same as those for the 
real vortex (additionally one needs to reverse the direction of the image vortex). A similar 
method was used by Zieve and Donev\cite{Zieve}. This image vortex method is accurate in 
case of straight vortices parallel to the cylinder axis. When comparing the simulations 
with the full solution of the Laplace equation, the agreement is quite good since only 
a minor part of the velocity comes from the image field.    

The vortex filament model cannot describe vortex reconnection processes and tell when 
a reconnection occurs. We use commonly accepted criteria for making a reconnection: 
A reconnection in simulations is done when two vortices approach each other closer 
than some small distance. This distance, denoted by $h_{\rm rec}$, is typically taken to be 
about one half of the smallest space resolution used to describe the vortices, $h_{\rm min}$. 
Additionally, we require that in every reconnection the vortex length must decrease, which 
approximately indicates that the total energy decreases. This additional requirement 
will prevent, for example, two parallel straight vortices from reconnecting. The distance
between vortices is calculated, not simply by determining the distance between the points,
but by calculating the smallest distance between different vortex segments. These
vortex segments (extending from ${\bf s}_i$ to ${\bf s}_{i+1}$ are assumed to
be straight. Reconnection with a cylinder wall is assumed to occur when a vortex is closer
than $h_{\rm rec}/2$ from the wall. For plane boundaries this means that the distance 
to the image vortex is smaller than $h_{\rm rec}$ used for bulk reconnections. When a 
reconnection is done, the point closest to the wall is split into two and both these points
(one being the first point on a vortex and the other one being the last point) are moved to 
the surface. The point separation along the vortex is kept between some reasonable limits, 
such that  $h_{\rm min} < \vert {\bf s}_{i+1}-{\bf s}_i\vert < h_{\rm max}$ and more points 
are used in regions where the vortex is strongly curved. At the boundaries the vortex 
tangent is fixed to be along the surface normal. This requirement ensures that 
there is no flow through the surface. Small vortex loops are removed when it becomes 
impossible to follow their dynamics. In our numerical scheme this is implemented by
removing any closed loop or vortex terminating on the wall which is shorter than $4h_{\rm min}$. 

\section{Results}\label{s.results}

\subsection{Configuration of a single vortex}

Before analysing how a vortex array behaves in the tilted cylinder it is informative to 
consider a single vortex. The stability of the vortex with respect to creation of new 
vortices\cite{precursorPRL} depends on mutual friction $\alpha(T)$ but also on the initial
configuration (which in many experiments is unknown). The lower the temperature, the more 
easily a vortex becomes unstable, leading to the creation of a new vortex after 
reconnection at the boundary. All this is consistent with experiments\cite{precursorPRL,nature}. 
However, with zero tilt angle $\eta$ only very few configurations lead to multiplication 
of vortices, see for example Refs.~\cite{simu,precursorPRL}. Therefore, to investigate the 
propagating vortex front\cite{frontPRL} and the twisted vortex state\cite{twistPRL} behind 
it, the simulations had to be started with a large number of vortices. 
At low temperatures with small mutual friction our simulations show that the probability 
for a vortex to become unstable increases as the tilt angle increases from zero. 
In order to observe easy vortex multiplication (for various initial configurations) 
the tilt angle should be much larger than the experimental one, which is
only of order one degree. 

\begin{figure}[tb]
\centerline{
\includegraphics[height=0.5\linewidth]{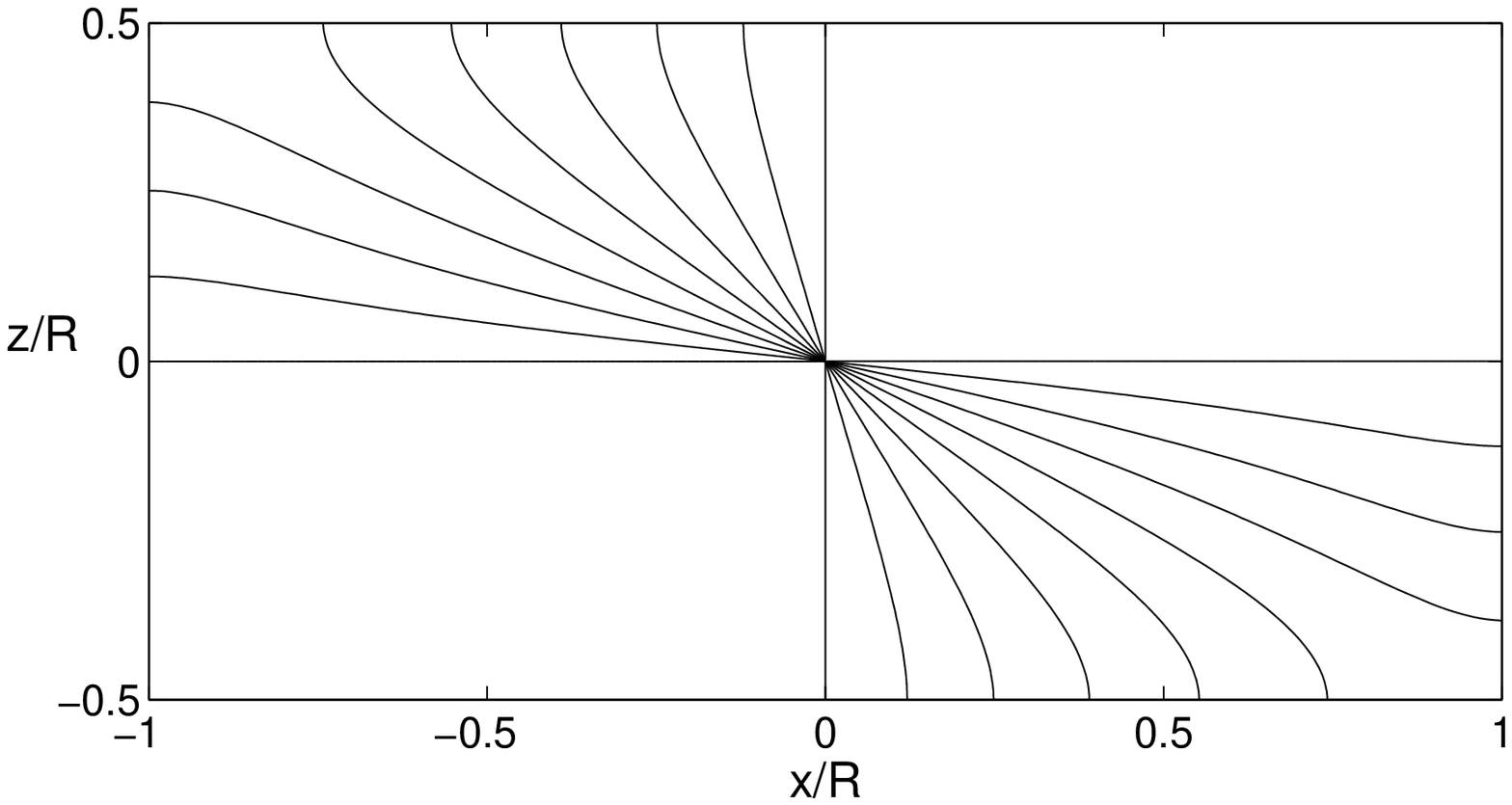} }
\centerline{
\includegraphics[height=0.91\linewidth]{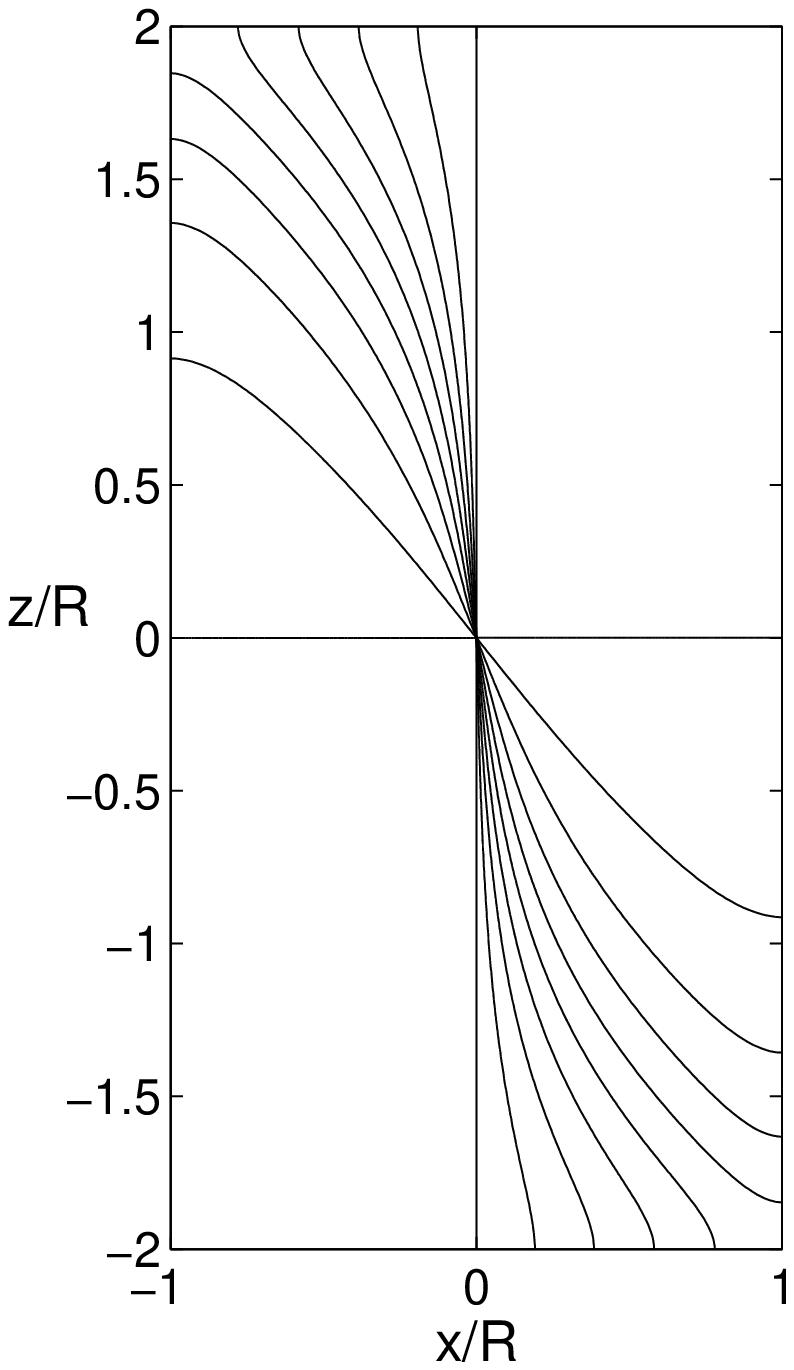}\hspace{0.15\linewidth}
\includegraphics[height=0.91\linewidth]{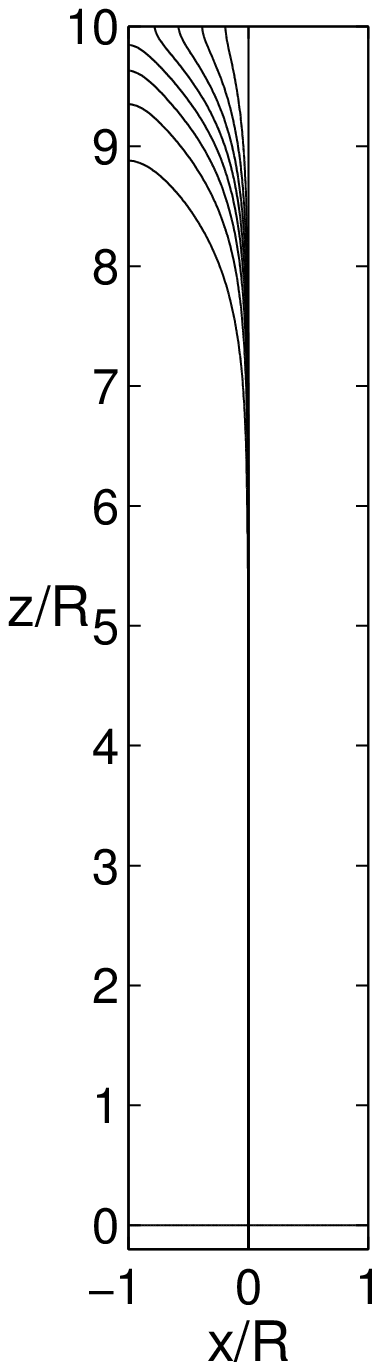}
}
\caption{Configuration of a single vortex in the tilted rotating
cylinder at a high rotation velocity of $\Omega = 500$ mrad/s and radius $R$ = 3 mm. 
Tilt angles are $\eta = 0, 10, 20, \ldots,$ and $90$ degrees. The vortex is practically 
in the $xz$-plane that is defined by the rotation and cylinder axes. {\it Top}: $L=R$. 
{\it Bottom left}: $L=4R$. {\it Bottom right}: $L=20R$, only the upper half of the
vortex configuration is plotted.}
\label{f.singlevor500mrads}
\end{figure}

In this article we are not focusing on the criteria for the vortex stability with respect
to reconnection on the wall and the formation of independently moving new loops. These
criteria seem to depend on many different features and parameters (at least on tilt angle, 
temperature, rotation velocity and initial vortex configuration). We are 
interested in the (possibly metastable) steady states in the cylinder, with 
different number of vortices in the cylinder. 

The easiest way to find the steady state configurations of a single vortex 
is to start with the vortex along the cylinder axis with zero tilt. Then by 
increasing the tilt angle (which was done in Fig.\ \ref{f.singlevor500mrads} 
in steps of 10 degrees) one may follow the vortex dynamics by iterating 
Eq.\ (\ref{e.vl}) until the vortex finds its equilibrium position. To speed 
up the process and to avoid vortex multiplication these simulations were 
done at high enough temperatures, with the $^3$He-B parameter values at $T=0.8T_c$.
The numerical resolution for these single vortex calculations was such that
$h_{\rm min}$ = 0.04 mm and $h_{\rm max}$ = 0.10 mm. The cylinder radius
was fixed to R = 3 mm.   

For large rotation velocities there was no indication of hysteresis, i.e. the
obtained configuration was the same even if one started the iteration from a straight 
vortex perpendicular the cylinder axis. The steady state configurations for high 
rotation velocity are illustrated in Fig.~\ref{f.singlevor500mrads}.
One may notice that for a single vortex the steady state depends strongly on the
aspect ratio $L/R$ of the cylinder. For short cylinders $L \sim R$  the vortex is 
approximately along the rotation axis, except near boundaries, where the vortex must 
be along the surface normal. For long cylinders, $L \gg R$, the vortex lies mainly 
along the cylinder axis and only deviates from this near the bottom and top plates. 
When $\eta \rightarrow 90^\circ$ the vortex finally turns and aligns itself along the 
rotation axis.  For long cylinders the vortex is along the cylinder axis because then the
azimuthal counterflow is reduced in a longer region, even if the reduction at every 
point is not an optimal one. For the short cylinder it is more important how well the 
counterflow is reduced locally and therefore a better result is obtained by orienting 
the vortex along the rotation axis. 
One should note that, with respect to the number of vortices, these single vortex 
states at high rotation velocities are metastable and far from equilibrium. However, 
they are experimentally accessible states in superfluid $^3$He-B, where 
at high temperatures one may easily create either a vortex free state or 
a state with an arbitrary number of vortices\cite{Ruutu1997}. 

\begin{figure}[tb]
\centerline{
\includegraphics[height=0.85\linewidth]{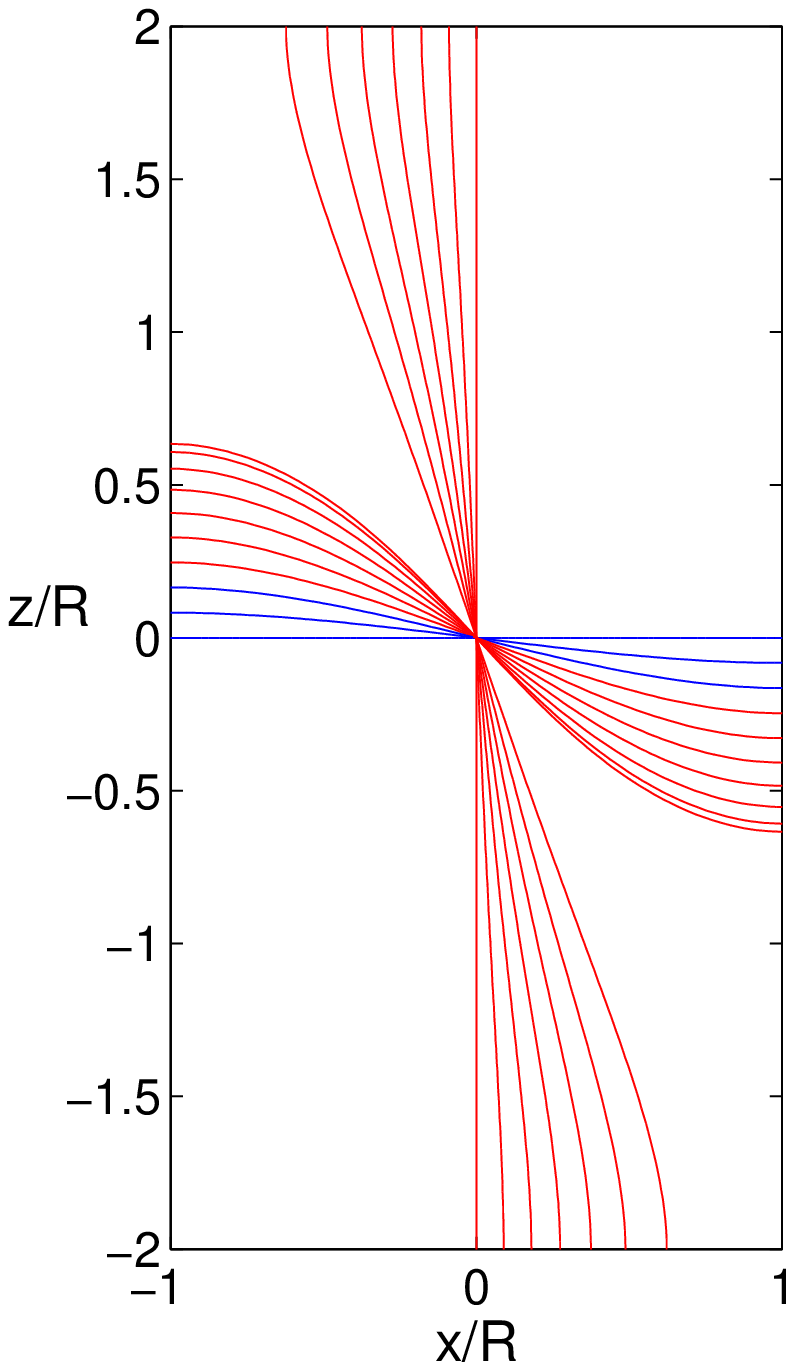}
\includegraphics[height=0.85\linewidth]{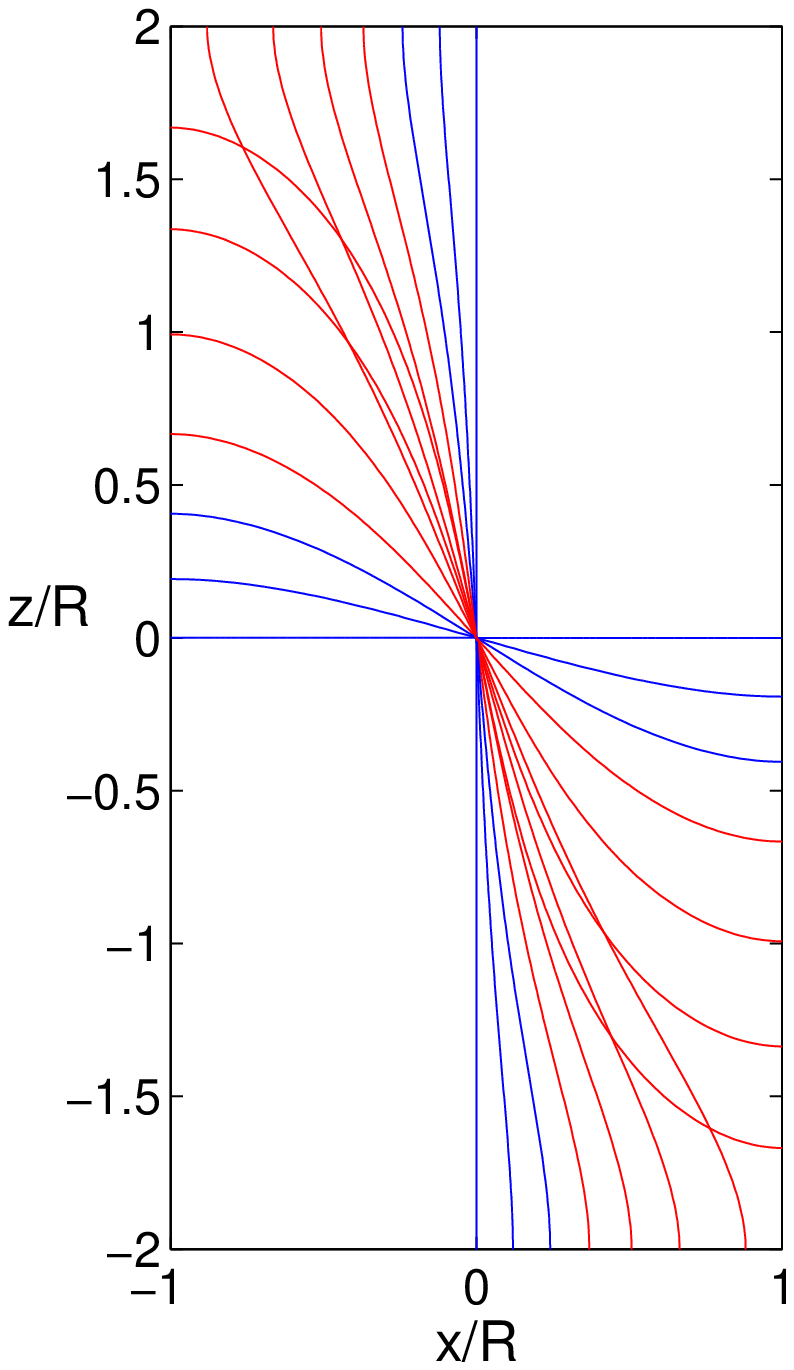}
}
\caption{(Color online) Steady state vortex configurations for a single vortex in
a tilted cylinder with $R$ = 3 mm and $L$ = 12 mm. The vortex is in the plane defined 
by the rotation and cylinder axes ($xz$-plane). 
{\it Left}: Low rotation velocity of 10 mrad/s.
The tilt angles for vortices terminating to the top and bottom ends of the cylinder are: 0, 10, 20, 
30, 40, 50, and 60 degrees and were obtained by increasing the tilt angle. The vortices terminating
to the outer cell wall ($\rho=R$) correspond to tilt angles of 0, 10, $\ldots$, 90 and were obtained by 
decreasing the tilt from 90 degrees. The configurations for the highest tilt angles (from 70 to 90)
can also be obtained by increasing the tilt. 
{\it Right}: Rotation velocity of 25 mrad/s. For vortices attached to the top and bottom plates of 
the cylinder the tilt angles are 0, 10, 20, $\ldots$, 60 degrees. For vortices terminating to the outer 
cell wall $\eta$ = 30, 40, $\ldots$, 90 degrees. Hysteresis in increasing/decreasing inclination 
is observed for angles 30, 40, 50, and 60 degrees. 
}
\label{f.singlevor2}
\end{figure}

At small rotation velocities one may observe hysteresis between different meta\-stable
configurations. This is quite natural since 
the boundary condition that a vortex must terminate perpendicular to the solid wall 
will result in a self-induced velocity due to its curvature that is of order 
$v_{\rm curv} = \kappa \ln(8R/a_0)/4\pi{R}$, where $R$ is the radius of the cylinder. 
This translates to a characteristic rotation velocity of order 10 mrad/s at which 
hysteresis becomes important. This is illustrated in Fig.~\ref{f.singlevor2} where 
we have followed the vortex configuration by increasing the tilt angle first from 
zero to 90 degrees and then decreased it back to zero (also in steps of 10 degrees). 
Hysteresis also results from the cell geometry and the boundary requirement at the 
solid wall. Therefore a vortex may get trapped either on the top/bottom plate or on 
the cylindrical side wall, $\rho=R$, in a similar fashion as a vortex is trapped on a 
spherical pinning site on the plane boundary\cite{schwarz85}.

\subsection{Vortex array in an infinitely long cylinder}

When the rotation and cylinder axes are perfectly aligned, the vortices form a lattice
of rectilinear lines which are parallel to the common axis. Closer to the cylindrical
outer wall the vortex array deforms to concentric rings. These are separated from the 
cylinder wall by a vortex-free annulus which has a thickness of order of the intervortex 
distance\cite{fetterSurf,donnellySurf,Ruutu1998}. If one considers an infinitely long 
cylinder, which naturally is impossible to organize experimentally, but which gives 
a good hint of what might occur in a long cylinder far from the ends, some analytical 
results are possible. In this case the potential velocity profile due to the tilt is 
obtained by putting $g=0$ in Eq.\ (\ref{e.velorot2}). One may easily note that an array 
of rectilinear vortices along the cylinder axis can still be a stable solution. However, 
now the areal vortex density $n_v = 2\Omega\cos\eta/\kappa$ is reduced by the factor 
$\cos\eta$ due to the reduced azimuthal counterflow in Eq.\ (\ref{e.velorot2}). The axial velocity $v_z$ in 
Eq.\ (\ref{e.velorot2}) does not affect the vortex configuration as long as it is 
below the critical value of the Ostermeier-Glaberson instability\cite{GJO,Donnelly}. 
Considering the stability of a single vortex and taking into account the maximum axial 
velocity $v_{\rm z,max} = 2R\Omega\sin\eta$ and the reduced counterflow, one obtains 
that the vortices suffer the instability when $v_{\rm z,max} >  2\sqrt{\nu\Omega\cos\eta}$, 
where $\nu = \kappa\ln(1/ka_0)/(4\pi) \approx \kappa$. This results in that the 
instability occurs when
\begin{equation}
\Omega > \Omega_{\rm c} = \frac{\nu\cos\eta}{R^2\sin^2\eta},
\label{e.Ocrit}
\end{equation}
with the wave number $k = \cot\eta/R$. In this derivation we assumed that the vortex
array fills the whole cylinder and we ignored the vortex free region near the walls. 
Alternatively, by considering a vortex cluster with $N$ vortices, with radius 
$R_{\rm c} = \sqrt{\kappa N/2\pi\Omega\cos\eta}$, placed in the center of the cylinder, 
one obtains that the vortices at the largest axial flow ($\rho = R_{\rm c}$ and 
$\sin\phi = \pm{1}$) suffer an instability when 
\begin{equation}
N > N_{\rm c} = 2\pi\frac{\nu}{\kappa}\cot^2\eta \approx 2\pi\cot^2\eta,
\label{e.Ncrit}
\end{equation}
which depends only on the tilt angle of the cylinder. This is rather well 
satisfied numerically if one additionally takes into account that for small 
$N$ the vortex bundle is always somewhat smaller than what the analytical 
estimate suggests. Using periodic boundaries along the cylinder axis and a 
period that is much larger than the possible wavelength of Kelvin waves, one obtains 
that with a tilt angle $\eta = 30^\circ$, the corresponding $N_{\rm c} = 19$ according
to Eq.\ (\ref{e.Ncrit}), while numerically we find that a bundle with 22 vortices is 
still stable. With 25 vortices one may observe growing Kelvin waves, first at locations 
where the axial velocity is largest. 

We have not determined the vortex configurations of an infinitely long cylinder 
in detail, since the computer code is not fully compatible with periodic
boundaries. However, some calculations have been done for finite cylinders, 
the case more appropriate experimentally. Some of these findings are 
reported below.

\subsection{Steady states in finite cylinder}

\begin{figure}[bt]
\centerline{\includegraphics[width=\linewidth]{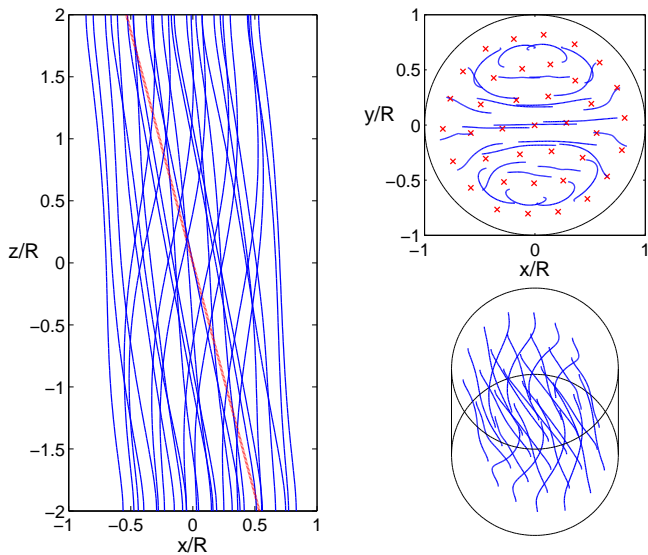}}
\caption{(Color online) Stable vortex array configuration, when $\Omega = 50$ mrad/s in 
a cylinder with $L = 4R = 12$ mm and a tilt angle of 15 degrees. 
{\it Left}: Side view along the $y$-axis where the red straight line denotes the rotation axis.
{\it Top right}: Top view where the crosses denote the equilibrium 
location of straight vortices at zero tilt, the configuration used as an 
initial guess for the calculations.. 
{\it Bottom right}: Perspective view of 
the same configuration. During the iteration 10 out of the 38 initial
vortices annihilate, which indicates that the counterflow at the cylindrical 
boundary needs to be large enough to prevent the vortices that terminate 
to the outer boundary from shrinking away. The calculations were done using 
mutual friction parameters that correspond to $T = 0.8T_{\rm c}$ for
$^3$He-B. A similar configuration can be obtained at lower temperatures 
where it only takes much longer time to reach equilibrium, owing to much smaller 
mutual friction. Numerical resolution is characterized by $h_{\rm min}$ = 0.04 mm and 
$h_{\rm max}$ = 0.25 mm. 
}
\label{f.arrayconf1}
\end{figure}

\begin{figure}[h]
\centerline{\includegraphics[width=\linewidth]{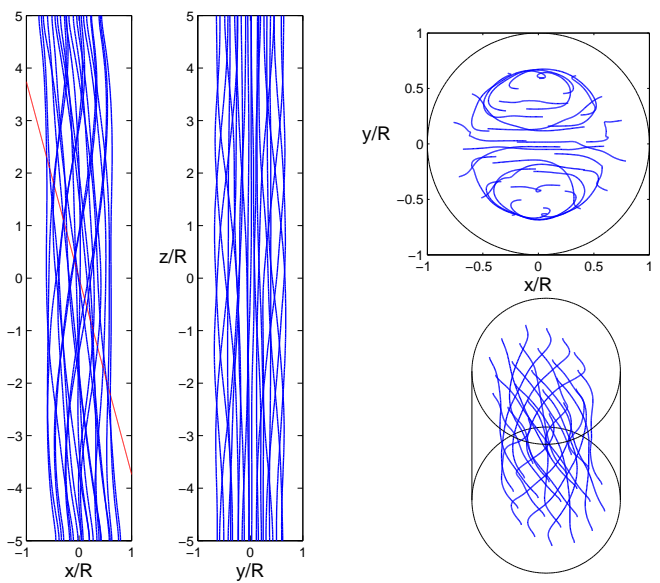}}
\caption{(Color online) Stable vortex array configuration, when $\Omega = 50$ mrad/s in 
a cylinder with $L = 10R = 30$ mm and a tilt angle of 15 degrees. 
{\it Left}: Side view along the $y$-axis where the red straight line 
denotes the rotation axis.
{\it Center}: Side view along the $x$-axis.
{\it Top right}: Top view. 
{\it Bottom right}: Perspective view of the same configuration. During the iteration 15 
out of the 38 initial vortices annihilate. The calculations were done using 
mutual friction parameters that correspond to $T = 0.8T_{\rm c}$ in $^3$He-B.
The initial configuration and the numerical resolution were the same as in Fig.~\ref{f.arrayconf1}.}
\label{f.arrayconf2}
\end{figure}

For small rotation velocities and small tilt angles one may 
observe a static steady state. This state is most likely not the absolute
energy minimum. Frequently the calculations for straight vortices have
shown that different initial configurations will lead to metastable
states with very similar energies\cite{fetterSurf}. Some of these steady 
states in the tilted cylinder are shown in Figs.~\ref{f.arrayconf1} and 
\ref{f.arrayconf2}. These configurations were obtained from an initial array
of straight vortices along the cylinder axis as the starting state and then 
iterating the equation of motion with a given tilt angle. As in the single vortex
case, one may notice that in a long cylinder the resulting vortex bundle is 
oriented more along the cylinder axis than along the rotation axis. Additionally, the
vortices become slightly twisted, which results from the tendency to reduce
the axial counterflow. This twist cannot be seen in the case of a single vortex, 
since the vortex lies in the plane where the axial flow is absent.


\begin{figure}[ht]
\centerline{\includegraphics[width=\linewidth]{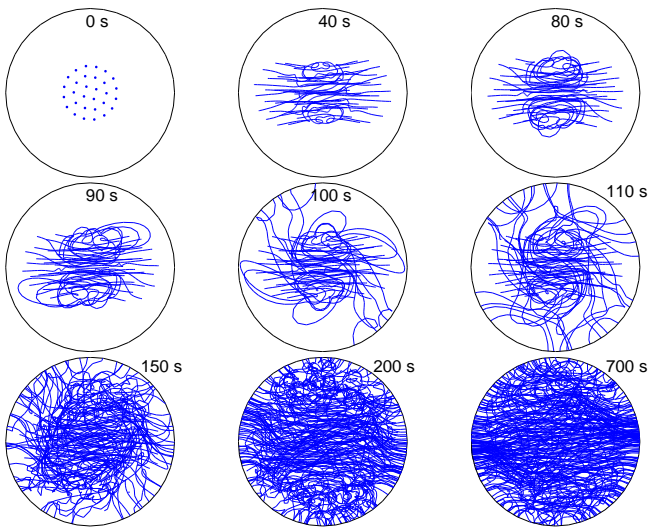}}
\caption{(Color online) Development of the dynamic steady state, when $L$ = $6R$ = 18 mm for
a rotation velocity of 250 mrad/s and a tilt angle of 30 degrees. The mutual friction parameters
correspond to $T = 0.4T_c$ in $³$He-B. The initial configuration is an array 
of 30 rectilinear vortices that become unstable due to the Glaberson instability. 
Similar break-up of the vortex array can also be seen in simulations done at $T =0.8T_c$. 
Numerical resolution is characterized by $h_{\rm min}$ = 0.10 mm and $h_{\rm max}$ = 0.30 mm.
See also Fig.~\ref{f.turb2}. 
}
\label{f.turb1}
\end{figure}

\begin{figure}[ht]
\centerline{\includegraphics[width=\linewidth]{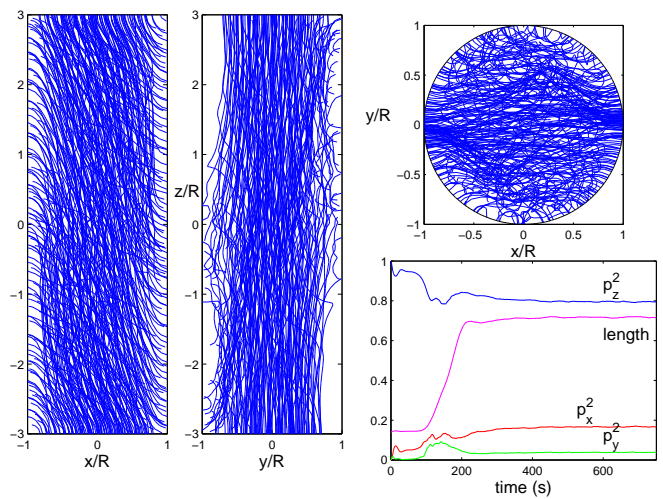}}
\caption{(Color online) Structure of the tangled bundle of vortices at 750 s in more detail for  
simulations with $L$ = $6R$ = 18 mm at a rotation velocity of 250 mrad/s and a tilt angle of
30 degrees. The mutual friction parameters correspond to $T = 0.4T_c$ in $³$He-B. 
{\it Left}: View along the $y$-axis that is perpendicular to the plane defined by the 
rotation and cylinder axes. 
{\it Center}: View along the $x$-axis. 
{\it Top right}: Top view along the cylinder axis.
{\it Bottom right}: Time development of the polarization 
$p_i^2 = \int(\hat{\bf s}'\cdot\hat{\bf i})^2d\xi/\int d\xi$ and the total length 
${\cal L} = \int d\xi$ of
the vortex tangle. This length ${\cal L}$ is scaled by the combined total length of rectilinear 
vortices in an array at zero tilt, in the continuum approximation that is 
$(2\Omega/\kappa)\pi{R}^2L$.  
}
\label{f.turb2}
\end{figure}

With large enough tilt and rotation velocity the initial array of straight
vortices suffers the Glaberson instability, as described above for an infinitely long
cylinder. Our simulations show that the equation for the critical number of vortices, 
Eq.~(\ref{e.Ncrit}), gives an estimate of the instability also in a
cylinder of finite length. If the number of vortices exceeds this value, then  
at the location of the largest axial flow the vortices go unstable. This is illustrated 
in Fig.\ \ref{f.turb1} where the simulations were started with 30 straight vortices 
at the center of the cylinder. The configuration at 750s is shown in Fig.\ \ref{f.turb2},
together with the time development of vortex length and average polarization,
$p_i^2 = \int(\hat{\bf s}'\cdot\hat{\bf i})^2d\xi/\int d\xi$. The initial vortex array, 
which has the polarization $p_z^2 = 1$ and $p_x^2 = p_y^2 = 0$, is broken by the 
Glaberson instability. This has the result that the polarization along the $z$-axis 
decreases at the same time as the vortex length increases. The final steady state 
value for $p_z^2 = 0.80$ is higher than what would result if all the vortices would 
be along the rotation axis, in which case  $p_z^2 = \cos^2\eta = 0.75$ 
(and $p_x^2 = \sin^2\eta = 0.25$). Since the turbulent fluctuations 
(Kelvin waves along vortices) and also the bending of vortices towards the radial 
direction near $\rho = R$ tend to decrease the polarization along the $z$-axis one
may safely argue that even in this dynamical steady state there is a tendency of the 
vortices to orient themselves along the symmetry axis of the cylinder.  

\begin{figure}[pht]
\centerline{\includegraphics[width=\linewidth]{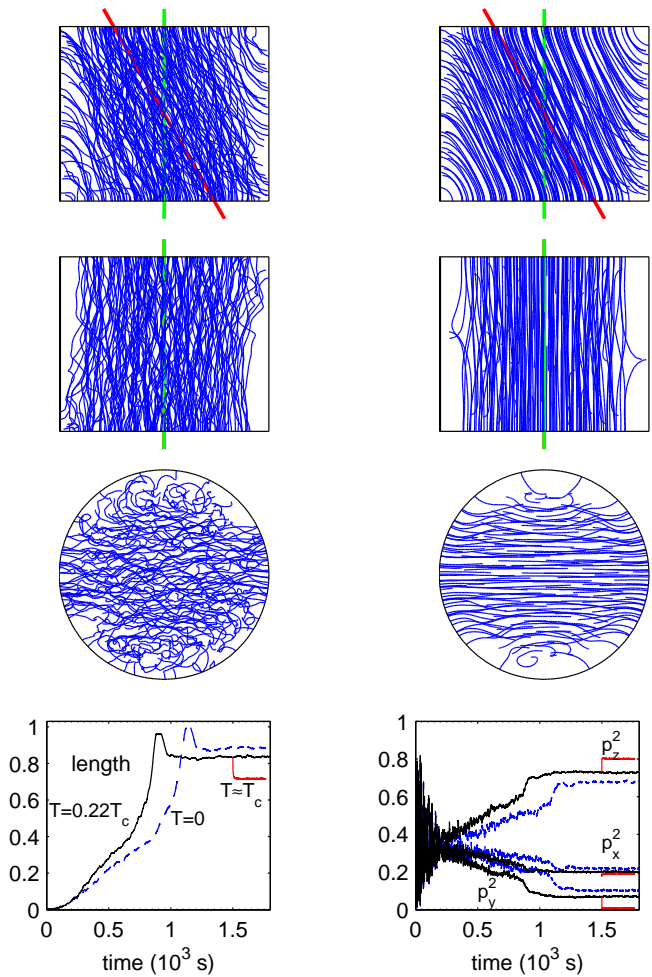}}
\caption{(Color online) Comparison between the low and high temperature vortex configurations 
for a short cylinder when $L$ = 5 mm, $R$ = 3 mm, $\Omega$ = 0.25 rad/s and $\eta$ = 30$^\circ$. 
Now the vortices are mostly along the rotation axis (noted by the red line in the uppermost plots). 
From top to bottom these three different views correspond to a view along the $y$-, $x$-, and $z$-axis, 
respectively. The low temperature configuration (taken at $t$ = 1800 s) on the left is for $T=0.22T_c$ 
in $^3$He-B and was obtained by using a single vortex, a quarter ring, as an initial configuration. 
The high temperature configuration on the right, which corresponds to mutual 
friction parameters $\alpha$ = 10 and $\alpha'$ = 0.9, was obtained by using the configuration 
at $t$ = 1500 s from the simulations at $T=0.22T_c$ as an initial guess and iterating for 
262 seconds.
{\it Bottom left}: Time development of the total vortex length, scaled by $(2\Omega/\kappa)\pi{R}^2L$.
The solid black line corresponds to $T=0.22T_c$ and the solid red line, which starts at $t$= 1500 s,
is the high temperature limit. The dashed blue line is a zero temperature calculation where
we have set $\alpha=\alpha'=0$ and the only dissipation is due to numerics. The initial 
configuration for the $T\rightarrow 0$ calculation was the same as for $T=0.22T_c$.  
{\it Bottom right}: Time development of the polarization of the vortex tangle. Color-coding 
is the same as for the vortex length. Large oscillations at small times result from the motion
of a single (or few) vortex around the cylinder.   
}
\label{f.turb3}
\end{figure}

A somewhat surprising result from this simulation (and some other similar ones which 
we have performed at somewhat different temperature or with different $L$) is that 
after the instability it is difficult to obtain a static steady state, 
which presumably would be the thermodynamic equilibrium configuration. Even after the 
vortex length and polarization of the vortex tangle have settled to a steady state 
value one may still observe creation of new vortices in regions of high axial counterflow 
(which arises from the potential flow) owing to the Glaberson instability and a 
reconnection with the cylinder boundary. These newly generated vortices are then driven 
towards less turbulent regions. At the same time there is continuous annihilation of 
small vortex loops near the cylinder ends in regions of small counterflow.
It is possible that our simulations do not extend far enough in time and that 
a static solution follows at much later times. Another possibility is that the 
calculation gets stuck in a local energy minimum and a possible static solution 
at lower energy may not be found with this initial configuration of vortices. This 
would also mean that experimentally it might be possible to observe a similar dynamic state. 

At high temperatures the vortices become smoother and the turbulent regions, that can be seen 
in Figs.~\ref{f.turb1} and \ref{f.turb2} at $T=0.4T_c$, become much more damped, but surprisingly some dynamics 
may still remain. For example, Kelvin waves are still created in the regions of largest axial flow. 
Similarly some vortices shrink in the regions of small counterflow near the cylinder ends. 
A comparison between the low and high temperature states is shown in Fig.\ \ref{f.turb3}. 
The low temperature state at $T=0.22T_c$ was obtained from an initial unstable configuration
with single vortex in the form of a quarter ring starting from the bottom of the cylinder and 
bending to the outer wall. At low temperatures with a tilt angle of 30$^\circ$ this configuration is 
unstable, leading to the creation of new vortices via reconnection at the boundary. At high 
temperatures the same initial configuration does not lead to vortex multiplication. Therefore the high 
temperature state, which here corresponds to $T\approx T_c$ (using $\alpha=10$ and $\alpha'=0.9$), 
was obtained using the state from the $T=0.22T_c$ calculation at $t$=1500 s as initial configuration 
and iterating the vortex motion with high mutual friction. The steady state which was obtained 
this way was still a dynamical one after 300s of evolution, but the creation and annihilation 
of vortices was a quite regular process. This can be seen, 
for example, by investigating the vortex length, which shows small fluctuations but also clear 
oscillations with a periodicity of order 10 seconds. Fig.\ \ref{f.turb3} shows also that the 
steady state vortex length increases and the axial polarization decreases with decreasing temperatures. 
This is due to fluctuations (Kelvin waves) on the vortices which are suppressed at high temperatures. 
The difference in steady state vortex length between the low and high temperature states is of order 20\%. 
In reality the difference is larger since simulations cannot describe structures that are smaller 
than the numerical resolution, which in these runs is determined by $h_{\rm min}$ = 0.10 mm and 
$h_{\rm max}$ = 0.30 mm. A small overshoot in vortex length, that can be seen in simulations at
low temperatures before the length saturates, indicates that a surplus of vortices is created 
before the final steady state is reached. Further analysis of the present simulations and more 
calculations are required to understand the complex dynamic phenomena in a tilted cylinder at 
large tilt angles with a large number of vortices. For example, the verification of Eqs.\ 
(\ref{e.Ocrit}) and (\ref{e.Ncrit}) is far from complete, especially when $\Omega_{\rm c}$ 
(or equally when $N_{\rm c}$) is large. Unfortunately the simulations take a lot of time and to 
determine whether the steady state is static or dynamical one is a tedious task.

\section{Conclusions}\label{s.conclusions}

We have investigated vortex motions in a rotating cylinder where the rotation 
axis is not parallel to the symmetry axis of the cylinder. We have determined
the steady state solutions and obtained that for small tilt angles the stable 
configuration consists of vortices that are along the cylinder axis, rather 
than the rotation axis. Therefore the assumptions behind the previously calculated 
result\cite{mathieu} might be inappropriate although there the calculations 
concerned a different flow geometry. For large tilt angles (but still well below 90$^\circ$) 
the steady state solution, especially at low temperatures, may be a dynamical one, 
which would make it possible to study a polarized turbulence in a well controlled 
environment, especially in superfluid $^3$He-B where strong vortex pinning can be ignored.
Since the vortex configuration is different from the expected one, it might be 
appropriate to reconsider the idea of a tilt-dependent mutual friction parameter, 
$B(\eta)$, as used in Ref.~\cite{mathieu} to interpret second-sound measurements 
in superfluid $^4$He. However, one should note that in He-II pinning is important 
and might alter the results.

\begin{acknowledgments}
I would like to thank M. Krusius for his valuable comments and improvements to the article.
This work is supported by the Academy of Finland (grant 114887). 
\end{acknowledgments}


\nopagebreak



\end{document}